\documentstyle[emulateapj,graphicx]{article}

\newcommand{\blank} {\lower 5pt\hbox to 0.75in{\hrulefill}}

\newcommand{\Osix} {O$\;${\footnotesize VI}$\;$}
\newcommand{\Ose} {O$\;${\footnotesize VII}$\;$/O$\;${\footnotesize VIII}$\;$}
\newcommand{\cgs} {erg~$\rm{s}^{-1}\rm{cm}^{-2}\rm{deg}^{-2}\;$}

\lefthead{Phillips, Ostriker, \& Cen}
\righthead{WHIM X-ray Background Spectrum}

\begin{document} 

\title{Is There Still Room for Warm/Hot Gas?
 Simulating the X-ray Background Spectrum}

\author{Lara Arielle Phillips, Jeremiah P. Ostriker and Renyue Cen}
\affil{Princeton University Observatory, Princeton, NJ 08544 \\
phillips, jpo, cen@astro.princeton.edu}

\begin{abstract}

At low redshifts, a census of the baryons in all known reservoirs
falls a factor of two to four below the total baryon density predicted from
Big Bang nucleosynthesis arguments
and observed light element ratios.
Recent cosmological hydrodynamic simulations suggest that 
a significant fraction of these missing baryons
could be in the form of warm/hot gas in the filaments
and halos within which most field galaxies are embedded.
With the release of source count results from Chandra
and recent detections of this gas in \Osix quasar absorption
lines, it becomes interesting to examine the 
predictions and limits placed on this component of the
X-ray background (XRB). 
We have used new hydrodynamical simulations to predict the total 
X-ray spectrum from the gas in the 100 eV to 10 keV range. 
We find that, when uncertainties
in the normalization of the observed XRB and the value of 
$\Omega_b$ are taken into account, our results are consistent with
current observational limits placed on the contribution
of emission from gas to the XRB. In the $0.5-2$ keV
range, we expect the contribution from this component to be 
$0.63 \times 10^{-12}$ \cgs or between $6\%$ and $18\%$ of the 
extragalactic surface brightness.
The peak fraction occurs in the $0.5-1$ keV range where 
the predicted line emission mirrors a 
spectral bump seen in the latest ASCA/ROSAT XRB data.

\end{abstract}


\section{Introduction}

Measurements 
of the total baryon density from the optical depth of the
Lyman $\alpha$ forest (Rauch et al. 1998) 
at high redshift ($z \approx 2-3$) are in remarkable agreement 
with those derived from observed light-element ratios and
nucleosynthesis arguments (BBN, Burles \& Tytler), which 
yield a value of $\Omega_{b,D/H} = 0.039 \pm 0.002$, assuming 
$h \equiv H_{0}/100 = 0.7.$ However, in the local universe, 
the combined observed contributions of stars, atomic
and molecular hydrogen, and cluster gas fall a factor of two to four 
below this number (Fukugita, Hogan, \& Peebles 1998). 
This is known as the missing baryon problem.

Cosmological hydrodynamical simulations suggest that, at low redshift,
a significant fraction, up to $40\%$, of all the baryons, could be in 
the warm/hot gas in filaments (e.g., Cen et al. 1995; Cen \& Ostriker
1999; Dav\'{e} et al. 2000; Croft et al. 2000).  
This diffuse gas, which is also known as the Warm/Hot Intergalactic
Medium (WHIM), is 
located in filamentary structures between 
clusters of galaxies and in which groups of
galaxies are embedded. It is shock-heated to
temperatures $T = 10^{5}-10^{7}$K, between that of the hot
cluster gas and the warm gas in voids. The 
comparatively low density and temperature of this gas make it a 
challenge to detect and as yet, attempts to observe it in emission have 
yielded marginal (Wang, Connolly \& Brunner 1997; Boughn 1998;
Kull \& B\"ohringer 1999; Scharf et al. 2000) or negative results, with an upper bound of 
$1.1 \times 10^{-12}$ \cgs  in the 0.5 to 2 keV band for intercluster gas 
(Briel \& Henry 1995).  However, warm gas has been detected as 
\Osix absorption features in quasar spectra 
(Tripp \& Savage 2000; Tripp, Savage, \& Jenkins 2000, and 
references therein)
 which seem to coincide in redshift 
with galaxy overdensities. 
The WHIM may have been indirectly detected in 
the observed X-ray background (XRB) autocorrelation  
(Soltan \& Freyberg 2000) and cross-correlation with the galaxy 
density distribution (Refregier, Helfand, \& McMahon 1997; Soltan et
al. 1997). 
In addition, rocket experiments show some evidence of 
\Ose emission from this gas (D. McCammon et al.,
in preparation).
  
Most of the soft XRB has now been resolved into AGN and 
other sources, leading to an upper bound on the fraction of the 
extragalactic soft $1-2$ keV X-ray component 
due to the WHIM of $25\%$ 
(Hasinger et al. 1998; Mushotzky et al. 2000; Giacconi et al. 2000). 
This is a small fraction of the XRB but could represent a large 
fraction of the baryons of moderate overdensity 
(Dav\'{e} et al. 2000).
 Markevitch (1999), Phillips, Ostriker \& Cen (2000)
 and Pierre, Bryan \& Gastaud (2000)
have already described possible methods to detect this gas 
using Chandra and XMM. In addition, 
the X-ray properties of this gas have been predicted from large scale 
simulations using both 
smoothed particle hydrodynamics (SPH) and 
Eulerian hydrodynamical  
methods (Croft et al. 2000; L. A. Phillips, in preparation).

We note that the hot gas component of the XRB is well
established. Hot gas associated with rich clusters of galaxies is
estimated to produce $\sim 10\%$ of the $1-2$ keV background 
(Hardy et al. 1998) and represents a fraction of approximately $8\%$
of the computed total baryonic complement of the cosmic 
mass density (Fukugita et al 1998). These diffuse 
baryons are associated with the $\sim 5\%$ of all galaxies (Bahcall 2000)
which are found in rich clusters. Most galaxies are in fact 
in lower density, lower velocity dispersion regions 
(such as poor clusters, groups, filaments and the field)
and would be
expected to be associated with lower density, lower temperature gas.
It is the purpose of this paper to ascertain if the expected emission
from this gas is consistent with our knowledge of the relatively soft
XRB.

We present the total 100 eV to 10 keV X-ray spectrum
from the WHIM calculated from new cosmological hydrodynamical simulations.
The simulation is described in Section 2 and 
discussed in the 
context of the current limits placed on the different 
components of the XRB, including point sources, cluster gas
and the more diffuse WHIM in Section 3. 
We summarize our results and briefly discuss the 
implied prospects of detecting the WHIM with ongoing and 
future X-ray missions in the Conclusions (Section 4).

\section{The Integrated Spectrum}

In the following calculations, we use the dark matter, galaxy, and
gas temperature, density and metallicity information
at redshifts zero, 0.5, 1 and 3 from the cosmological hydrodynamical 
simulation of a $L=100h^{-1}$Mpc box, with $512^{3}$ fluid elements and
$256^{3}$ dark matter particles, described in detail in Cen \&
Ostriker (1999). The simulation 
follows the metallicity history of the gas in each cell, 
enabling us to examine the contribution of metal lines 
to the XRB spectrum. The helium mass fraction is set to
that of primordial gas, Y $= 0.24$. 
The chosen cosmology is the "concordance model" (Wang et al.
2000), 
a flat low-density ($\Omega_{0} = 0.37$) universe with a Cosmological
Constant ($\Omega_{\Lambda} = 0.63$), $\Omega_{b} = 0.035$ and $H_{0}
= 70.$
The AGN integrated spectrum from a
 new, higher resolution cosmological hydrodynamical simulation 
of a smaller $L=25h^{-1}$Mpc box with $768^{3}$ fluid elements 
(R. Cen \& J. P. Ostriker, in preparation) is also used in 
estimating the large box AGN XRB spectrum.

We assume that the gas is in thermal and ionization
equilibrium, and use the Raymond-Smith (1977) code 
for optically thin plasmas, as modified by 
Cen et al. (1995), to obtain
the average Bremsstrahlung and 
emission line spectrum from the
large box for each of the four output redshifts.
We then integrate the $z= 0, 0.5, 1,$ and $3$
spectra, assuming no evolution, over the redshift ranges
bounded by redshifts   
$z_{bound} = 0, 0.25, 0.75, 2$ and $3$, 
to obtain the total integrated spectrum from the WHIM.

We remove the contribution from
gas within $1.5 h^{-1}$ Mpc of the center of
low redshift ($z \leq 0.2$) clusters in the simulation 
which were associated with (dark matter + baryonic) 
masses $M \geq 5.5 \times
10^{14} h^{-1} M_\odot$ (or richness $\geq 2$, Bahcall \& Cen 1993)
by Nagamine, Cen, \& Ostriker (2000). 
Cluster gas emission regions are usually treated as ``sources''
and removed from the total observed background when attempting to
estimate the residual ``hot gas background''. The component we are
addressing in this paper is of course not uniformly distributed. It is
also clumped, albeit with a far lower clumping factor than the hot gas in
clusters (see Dav\'e et al. 2000 for a detailed discussion). We
nevertheless remove the cluster gas from our computation to make this
analysis comparable
with those usually undertaken in studies of the X-ray sky.

A composite AGN spectrum was obtained from the small box
by integrating an AGN template spectrum (Edelson \& Malkan 1986)
over all redshifts,
adopting an AGN emission efficiency
in keeping with those observed and proportional to the star 
formation rate averaged over the whole box (Cen et al. 1998).
We use this to estimate the contribution to the XRB spectrum 
from AGN in the large box.
We normalize the integrated AGN spectrum so that
 at high energies ($> 30$ keV) the XRB from the large
box is completely resolved into AGN (Mushotzky et al. 2000).
Cen \& Ostriker (1999)  
integrated the XRB spectrum 
from AGN, Orion-like stars, and Bremsstrahlung emission from gas
over all redshifts for the large box. 
We can recover the $0.1-10$ keV portion of this spectrum
to within a few percent
by adding the WHIM Bremsstrahlung spectrum 
obtained above to the estimated contribution from the AGN.
This sum will therefore be used as the total
simulated XRB spectrum in the discussion below.


The total integrated X-ray spectrum from the WHIM 
(dotted line) is shown in the upper panel
of Figure 1 along with the total XRB spectrum from 
the large box simulation (solid line). The latest
XRB spectra (T. Miyaji et al. 2000, private communication)
obtained from the ASCA LSS
region for the $0.6 - 10$ keV range (filled circles) and from 
ROSAT PSPC measurements 
at lower energies (filled triangles) are also shown. 
The simulation results scale as $\Omega_b^{2}$ and our 
uncertainty concerning this basic parameter 
must be added to the other uncertainties. 
The errors due to the finite number of simulation output redshifts 
and that introduced by only integrating out to a redshift of $z = 3$
are both negligible when compared with this uncertainty. 
Average published values of $\Omega_b h^{2}$ range from the low BBN value 
of 0.019 (Burles \& Tytler 1998) to a high estimate of 0.035 from
BOOMERANG data (Lange et al. 2000).
Ultimately, the uncertainty with regard to simulated surface 
brightness is not less than a factor of three.
The predicted fraction due to the WHIM is shown 
in the lower panel as a function of the  
observed ASCA and ROSAT backgrounds (symbols) and as a function
of the total integrated XRB from our simulations 
(dotted line). 

\section{The Components of the X-ray Background}

Deep X-ray images (Hasinger et al. 1998; Mushotzky et al. 2000; 
Giacconi et al. 2000) have shown that at least 
75\% of the soft XRB is accounted for by sources
(including X-ray emission from hot gas in clusters).
The percentage of the background remaining 
after currently known contributions from AGN and other resolved
sources have been subtracted is shown in the lower panel of Figure 1, 
for extremum values of the XRB. 
A sizeable $25 \pm 6\%$ of the latest 
measurement of the $1 - 2$ keV XRB from the ASCA LSS field,
$4.5 \pm 0.33 \times 10^{-12}$ \cgs
(T. Miyaji et al. 2000, private communication), remains unresolved.
Previous combined ROSAT and ASCA measurements of the XRB have yielded similar
or lower average surface brightnesses, as is evidenced by the
Barcons, Mateos \& Caballos (2000) Bayesian fit 
to XRB measurements at 1 keV of 
$6.63^{+0.4}_{-0.6}\times 10^{-26}$ 
erg~$\rm{s}^{-1}\rm{cm}^{-2}\rm{sr}^{-1}\rm{Hz}^{-1}$  
(with $90\%$ confidence errors),
which appears as the square in the upper panel of Figure 1. 
For the low value for the XRB of 
$3.7 \pm 0.53 \times 10^{-12}$ \cgs
(Gendreau et al. 1995) the remainder becomes $9 \pm 13\%$, 
consistent with zero.
 Taking their upper limit from
Chen, Fabian \& Gendreau (1997), Mushotzky et al. (2000) conclude that at most
$10^{-12}$ \cgs of the soft XRB 
remains unresolved. This value is consistent our prediction 
of 
$0.22 \times 10^{-12}$ \cgs from the WHIM 
(or $9\%$ of the computed total) in the same energy band.
When possible, we have estimated the error bars 
from information given in the above papers and by assuming that the error
for the Hasinger et al. (1998) counts is comparable to that of 
Mushotzky et al. (2000). This has most probably led to an underestimate 
of the errors involved in the other quantities described.

In the $0.5-2$ keV band the predicted contribution of emission from
gas to the total background is $\sim 13\%$.
At first glance, this value is quite a bit lower than   
the $35\%$ from the intergalactic medium 
obtained by Croft et al. (2000).
However this discrepancy can be explained by the inclusion in the 
Cen \& Ostriker (1999) code of substantial feedback and our excluding
 the contribution from nearby clusters. 
We obtain an AGN contribution of $86\%$. The remaining $1\%$ 
of the emission
comes from hot gas in nearby rich clusters and agrees remarkably 
well with the observed $1.4\%$ associated with ($z \leq 0.2$) Abell 
(1958) clusters (Soltan et al. 1996).
As expected, in the $2-10$ keV band, the WHIM contribution falls to
$5\%$ of the computed background. 
 Metal line emission produces $49\%$ ($11\%$) of the $0.5-2$ keV 
($2-10$ keV) WHIM contribution. The peak fraction occurs in the 
$0.5-1$ keV range where a spectral bump coincides with that seen in
the ASCA and ROSAT data.
The factor of two difference between the normalization of our 
total computed background 
and that of the observed background
may be due to our low assumed value of $\Omega_{b}$ or to an
underestimate of the observed AGN formation rates which we used to compute
the AGN spectrum. 
If we assume that the latter is indeed the case, we can compare
the contribution from gas predicted by our simulations with the observed
background. In the $1 - 2$ keV range, this results in a $6\%$ contribution
from the WHIM (up to $18\%$ using the upper limit value of $\Omega_{b}$).

If one accepts the latest value of the
XRB as correct, then a portion of the
soft XRB has yet to be explained. 
It is unlikely that all of the remainder will be resolved into AGN, 
since fainter AGN have harder spectra and the high spatial resolution 
(0.5 arcsec) of Chandra ensures that most of the sources 
have now been resolved. 
We know that most of the XRB is due to sources but there
must also be a contribution from the gas whose presence has been 
established in clusters, groups and galaxies.
In addition, the amount attributable to gas emission can
only increase 
as flux and surface brightness observational limits are improved
and the WHIM at
the outskirts of groups and clusters is detected.
It is therefore likely that the remaining portion of the 
XRB is due to a combination of gas and some (point) sources.

The current detections of and limits on the WHIM emission 
in the $0.5 - 2$ keV band can be compared with
the integrated surface brightness
of $ 0.63 \times 10^{-12}$ \cgs which we obtain
from the simulated XRB gas spectrum.
(This value is comparable to that quoted in Croft et al. (2000), if
one takes into account a factor of 1.8 difference in $\Omega_{b}^{2}$).  
Briel \& Henry (1995) used ROSAT All-Sky Survey data to search 
for filaments between 40 Abell (1958) cluster pairs 
separated by less than $25 h^{-1}$ Mpc.
They placed a $2 \sigma$ upper limit on the filament X-ray 
surface brightness of  $1.1 \times 10^{-12}$ \cgs in the 
$0.5-2$ keV band, using a plasma temperature of 0.5 keV, typical 
of those seen for the intercluster medium
in the hydrodynamical simulations quoted above.
 
In a deep ROSAT PSPC observation 
around the galaxy cluster A2125, at $z=0.25$, 
Wang, Connolly \& Brunner (1997) 
find evidence of low surface brightness X-ray emission 
from $0.85$ keV gas with an extent of 
$\sim 1.7 h^{-1}$ Mpc (11 arcmin). It
appears correlated with an overdensity of galaxies and is 
probably part of 
a $\sim 6 h^{-1}$ Mpc (35 arcmin) superstructure which 
they liken to filamentary structures seen in simulations.
Kull \& B\"{o}hringer (1999) found evidence of X-ray emission
extending over $\sim 9 h^{-1}$ Mpc,
connecting three clusters in combined ROSAT PSPC and ROSAT
All-Sky Survey observations of the core of the Shapley Supercluster.
Both of these detections yield surface brightnesses that are a factor
of 4 to 10 higher than the upper limit established by 
Briel \& Henry (1995)
and may be evidence of gas stripped during cluster interactions
rather than true filament gas. However, the gas temperature and physical
extent of the emission are consistent with the filamentary 
structures seen in our simulations. 
In their analysis of a deep ROSAT PSPC field, Scharf et al. (2000) 
find a half degree filamentary structure
with a $0.5-2$ keV surface brightness of $0.58 \times 10^{-12}$ \cgs.  
This filament is also present in the form of
an I Band galaxy overdensity.
The X-ray surface brightness of this $5 \sigma$ significance detection
 is well within the residual limits outlined above
 and is consistent with the computed average surface brightness.

\section{Conclusions}

Our prediction for the fractional X-ray contribution to the 
total XRB spectrum is consistent with the latest limits
from AGN and other resolved sources in deep pointings from ROSAT 
(Hasinger et al. 1998) and Chandra (Mushotzky et al. 2000; Giacconi
et al. 2000). 
The average integrated surface
brightness from the gas of $0.63 \times 10^{-12}$ \cgs
 in the 0.5 - 2 keV band falls below the upper 
limits set by Briel \& Henry (1995) and therefore, as of yet, 
the presence of gas between pairs of clusters cannot be ruled out.
We are just reaching the flux limits necessary for the detection of
this gas. The expected average flux levels lead us 
to predict that the WHIM could be observable
with current missions such as Chandra and XMM.  
This possibility is explored in greater detail for XMM in 
Pierre et al. (2000) 
and for Chandra in Phillips et al. (2000). 
In the next decade, an effort 
should be made to quantify this WHIM component
 in clusters, groups, galaxies,
filaments and/or sheets since a significant fraction of the 
baryons in the universe could be in this medium.

One of the major challenges in detecting the intermediate temperature 
gas in filaments will be to confirm the extragalactic origin of emission
whose broad band spectral signature matches that predicted for 
the WHIM.
The cross-correlation of the observed XRB and projected galaxy 
density distribution proves to be an ideal tool for this (L. A. Phillips,
in preparation). The cross-correlation (CCF) flattens into a signature plateau at 
at scales of a few to ten arcmin
(Refregier et al. 1997; Soltan et al. 1997). 
Soltan \& Freyberg (2000) also see this feature in XRB
autocorrelation calculations. The plateau in the CCF can be reproduced by 
the correlation
of the X-ray emission from the WHIM in filaments with the simulated 
projected galaxy distribution (Phillips et al. 1999; Croft et al.
2000; L. A. Phillips, in preparation).
If the gas is not detected at the predicted fluxes, 
the metallicity history of the WHIM in the simulations
should be closely re-examined since a large fraction of
the soft X-ray emission from this gas arises from metal line emission.
Indeed, the peak fractional contribution from the WHIM occurs in the
$0.5-1$ keV range where we predict a bump due to line emission
coincident with a similar bump seen in the observed ASCA
and ROSAT XRB spectrum.

The presence of the WHIM in groups of galaxies has already been 
established (Mulchaey et al. 1996). Groups that reside in galaxy filaments tend to have 
higher diffuse X-ray fluxes, indicating a higher density of gas 
(Mahdavi et al. 2000). 
Warm gas has also been detected in the halos of galaxies and in galaxy
filaments (both in \Osix absorption and in X-ray emission, 
see references above).
Therefore the question is not whether there is a WHIM, but rather 
how much of it there is and where it is located. And once the 
answer to this question is known, we can move towards a better 
understanding of the physical state of most of the
baryons in the universe.

\acknowledgments

We thank Ken Nagamine, Todd Tripp, Michael Strauss and Romeel Dav\'{e}
for fruitful discussions, as well as Takamitsu Miyaji for the
use of his X-ray background spectrum data.
This research was supported by NSF grants ASC-9740300 and AST-9803137.
LAP was supported in part by an award from the NSERC (Canada) 
and a grant from Zonta International.

\clearpage

\begin{figure}
\begin{center}
\scalebox{0.75}[0.75]{\includegraphics{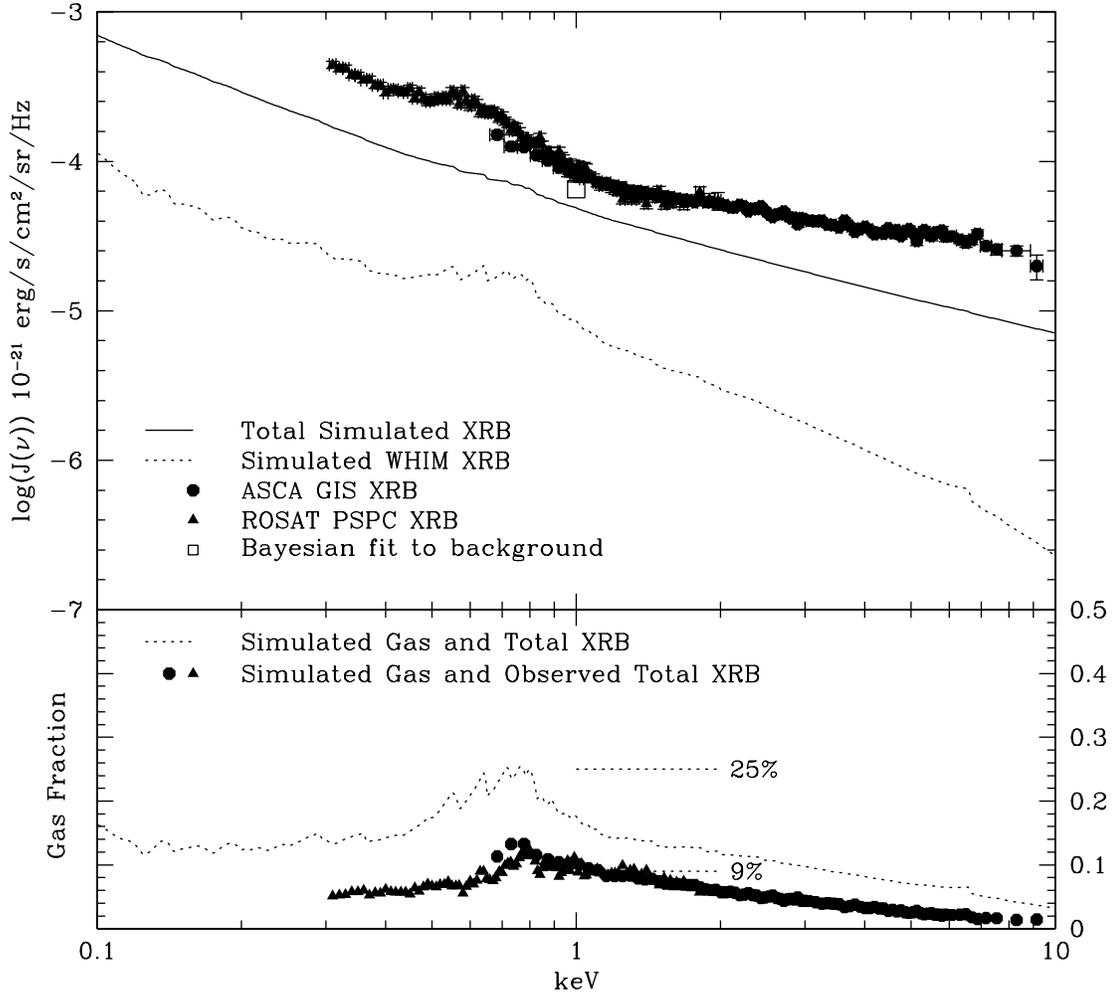}}
\end{center}
\caption{Integrated XRB from AGN, stars and 
gas (solid line) and from the WHIM alone (dotted line). 
The latest Miyaji et al. (2000, private communication) 
observations of the XRB spectrum from the ASCA LSS region 
and ROSAT PSPC fields (circles and triangles, respectively)
are also shown along with the
Barcons et al. (2000) Bayesian fit to observations of the X-ray
background (square, vertical scale illustrates size of error-bars).
The fractional contribution of the WHIM to the simulated total (line) 
and Miyaji et al. (2000) observed backgrounds (symbols) 
is plotted in the lower panel.
\label{fig1}}
\end{figure}


\begin{references}
\reference{abe58} Abell, G. O. 1958, ApJ Suppl., 3, 211
\reference{bah00} Bahcall, N. A. 2000, in 
\emph{Astrophysical Quantities}, ed. A. Cox (New York: AIP)
\reference{bah93} Bahcall, N. A., \& Cen, R. 1993, ApJ, 407, L49 
\reference{bar00} Barcons, X., 
Mateos, S., \& Ceballos, M. T. 2000, MNRAS, 361, 13
\reference{bou98} Boughn, S. P. 1999, ApJ, 526, 14
\reference{bri95} Briel, U.~G., \& Henry, J.~P. 
1995, A\&A, 302, L9
\reference{bur98} Burles, S., \& Tytler, D. 
1998, ApJ, 499, 699
\reference{cen95} Cen, R., Kang, H., Ostriker, J. P., 
\& Ryu, D. 1995, ApJ, 451, 436 
\reference{cen99} Cen, R., \& Ostriker, J.~P. 
1999, ApJ, 514, 1
\reference{cen98} Cen, R., Phelps, S.,
Miralda-Escud\'{e}, J., \& Ostriker, J. P. 1998, ApJ, 496, 577
\reference{che97} Chen, L.-W., 
Fabian, A. C., \& Gendreau, K. C. 1997, MNRAS, 285, 449
\reference{cro00} Croft, R. A. C., Di Matteo, T., 
Dav\'{e}, R., Hernquist, L., Katz, N., Fardal, M., \& 
Weinberg, D. H. 2000, ApJ, submitted, astro-ph/0010345
\reference{dav00} Dav\'{e}, R., et al. 2000, 
ApJ, submitted, astro-ph/0007217
\reference{ede86} Edelson, R. A., \& Malkan, M.
A. 1986, ApJ, 308, 59
\reference{fuk98} Fukugita, M., 
Hogan, C.~J., \& Peebles, P.~J.~E. 1998, ApJ, 503, 518
\reference{gen95} Gendreau, K. C. et al.
1995, PASJ, 47, L5
\reference{gia00} Giacconi, R., et al. 2000, 
ApJ, submitted, astro-ph/0007240
\reference{har98} Hardy, I. M., et al. 1998, MNRAS,
295, 641
\reference{has98} Hasinger, G., Burg, R., 
Giacconi, R., Schmidt, M., Tr\"{u}mper,
J., \& Zamorani, G. 1998, A\&A, 329, 482
\reference{kul99} Kull, A., \& 
B\"{o}hringer, H. 1999, A\&A, 341, 23
\reference{lan00} Lange et al. 2000, PRD, submitted
\reference{mah00} Mahdavi, A., B\"ohringer, H., 
Geller, M., \& Ramella, M. 2000, ApJ, 534, 114  
\reference{mar99} Markevitch, M. 1999, ApJ, 522, L13
\reference{miy96} Miyaji, T., Ishisaki, Y., 
Ogasaka, Y., Ueda, Y., Freyberg, M. J.,
 Hasinger, G., \& Tanaka, Y. 1998, A\&A, 334, L13
\reference{mul96} Mulchaey, J. S., Mushotzky, R. F., 
Burstein, D. \& Davis, D. S. 1996, ApJ, 456, L5 
\reference{mus00} Mushotzky, R.~F., Cowie, L.~L., Barger, A.~J., \& Arnaud, K.~A. 2000,
Nature, 404, 459
\reference{nag00} Nagamine, K., 
Cen, R., \& Ostriker, J.~P. 2000, ApJ, 541, 25
\reference{phi99} Phillips, L.~A., 
Ostriker, J.~P., Freyberg, M.~J., \& Tr\"{u}mper, J.
1999, BAAS, 195, 13.03 
\reference{phi00} Phillips, L.~A., 
Ostriker, J.~P., \& Cen, R. 2000, 
in the Proc. of "X-ray Astronomy 2000", eds. R. Giacconi, L. Stella
\& S. Serio, ASP Conf. Ser., in press
\reference{pie00} Pierre, M., 
Bryan, G., \& Gastaud, R. 2000, A\&A, 356, 403 
\reference{rau98} Rauch, M., et al. 1998, ApJ, 489, 1
\reference{ray77} Raymond, J.~C., \& 
Smith, B.~W. 1977, ApJS, 35, 419 
\reference{ref97} Refregier, A.~, 
Helfand, D., \& McMahon, R.~G. 1997, ApJ, 479, L93
\reference{sch00} Scharf, C., Donahue, M., 
Voit, G. M., Rosati, P., \& Postman, M. 2000, 528, L73
\reference{sol00} Soltan, A.~M.~, \& Freyberg, M.~J.
2000, in the Proc. of "X-Ray Astronomy 2000", eds. 
R. Giacconi, L. Stella \& S. Serio, ASP Conf. Ser, in press
\reference{sol96} Soltan, A.~M.~, Hasinger, G.~, 
Egger, R., Snowden, S., \& Tr\"{u}mper, J. 1996, A\&A, 305, 17
\reference{sol97} Soltan, A.~M.~, Hasinger, G.~, 
Egger, R., Snowden, S., \& Tr\"{u}mper, J. 1997, A\&A, 320, 705
\reference{tri00a} Tripp, T. M., \& Savage, B. D. 2000, ApJ, 542, 42
\reference{tri00b} Tripp, T. M., 
Savage, B. D., \& Jenkins, E. B. 2000, ApJ, 534, L1
\reference{van00} Vanden Berk, D. E., 
Stoughton, C., Crotts, A. P. S., Tytler, D., \& Kirkman, D. 2000, 
AJ, 119, 2571
\reference{wan00} Wang, L., Caldwell, R. R., 
Ostriker, J. P., \& Steinhardt, P. J. 2000, ApJ, 530, 17
\reference{wan97} 
Wang, Q. D., Connolly, A.~J., \& Brunner, R.~J. 1997, ApJ, 487, L13 
\end{references}
\end{document}